\documentclass[10pt,journal]{IEEEtran}
\usepackage{amsmath,amssymb,amsfonts}
\usepackage{algorithmic}
\usepackage{algorithm}
\usepackage{array}
\usepackage[caption=false,font=normalsize,labelfont=sf,textfont=sf]{subfig}
\usepackage{textcomp}
\usepackage{stfloats}
\usepackage{url,hyperref}
\hypersetup{
    colorlinks=true,     
    urlcolor=blue,
    }
\usepackage{verbatim}
\usepackage{graphicx}
\usepackage{gensymb}
\usepackage{cite}
\usepackage{xcolor}
\usepackage{multirow}
\usepackage{booktabs}
\usepackage[a4paper, total={184mm,239mm}]{geometry}
\usepackage[para,online,flushleft]{threeparttable}
\usepackage[detect-none,load-configurations=binary]{siunitx}
\DeclareSIUnit{\nothing}{\relax}
\def\BibTeX{{\rm B\kern-.05em{\sc i\kern-.025em b}\kern-.08em
    T\kern-.1667em\lower.7ex\hbox{E}\kern-.125emX}}

\begin{document}

\title{
Accelerating Image-based Pest Detection on a Heterogeneous Multi-core Microcontroller  
}

\author{
\IEEEauthorblockN{
Luca Bompani\IEEEauthorrefmark{1},
Luca Crupi\IEEEauthorrefmark{2},
Daniele Palossi \IEEEauthorrefmark{2}\IEEEauthorrefmark{3},
Olmo Baldoni\IEEEauthorrefmark{1}, \\
Davide Brunelli\IEEEauthorrefmark{4},
Francesco Conti\IEEEauthorrefmark{1},
Manuele Rusci\IEEEauthorrefmark{5},
Luca Benini \IEEEauthorrefmark{1}\IEEEauthorrefmark{3}
}  

\IEEEauthorblockA{\IEEEauthorrefmark{1} Department of Electrical, Electronic and Information Engineering, University of Bologna, Italy}

\IEEEauthorblockA{\IEEEauthorrefmark{2} Dalle Molle Institute for Artificial Intelligence, USI-SUPSI, Switzerland}

\IEEEauthorblockA{\IEEEauthorrefmark{3} Integrated Systems Laboratory, ETH Z\"urich, Switzerland}

\IEEEauthorblockA{\IEEEauthorrefmark{4} Department of Industrial Engineering, University of Trento, Italy}

\IEEEauthorblockA{\IEEEauthorrefmark{5} Department of Electrical Engineering, KU Leuven, Belgium}

Contact author: luca.bompani5@unibo.it
}

\IEEEpubid{ \hspace{-9.5cm}\begin{minipage}{\columnwidth} \vspace{0.8cm}
978-1-6654-XXXX-X/24/\$31.00-2024 IEEE. Personal use of this material is permitted. Permission from IEEE must be obtained for all other uses, in any current or future media, including reprinting/republishing this material for advertising or promotional purposes, creating new collective works, for resale or redistribution to servers or lists, or reuse of any copyrighted component of this work in other works.
\end{minipage}}

\maketitle
\IEEEpubidadjcol

\begin{abstract}
The codling moth pest poses a significant threat to global crop production, with potential losses of up to 80\% in apple orchards.
Special camera-based sensor nodes are deployed in the field to record and transmit images of trapped insects to monitor the presence of the pest. 
This paper investigates the embedding of computer vision algorithms in the sensor node using a novel State-of-the-Art Microcontroller Unit (MCU), the GreenWaves Technologies' GAP9 System-on-Chip, which combines 10 RISC-V general purposes cores with a convolution hardware accelerator. 
We compare the performance of a lightweight Viola-Jones detector algorithm with a Convolutional Neural Network (CNN), MobileNetV3-SSDLite, trained for the pest detection task. 
{
On two datasets that differentiate for the distance between the camera sensor and the pest targets, the CNN generalizes better than the other method and achieves a detection accuracy between 83\% and 72\%.
Thanks to the GAP9's CNN accelerator, the CNN inference task takes only \SI{147}{\milli\second} to process a 320$\times$240 image. Compared to the GAP8 MCU, which only relies on general-purpose cores for processing,  we achieved 9.5$\times$ faster inference speed.}
{When running on a  \SI{1000}{\milli\ampere\hour} battery at \SI{3.7}{\volt}, the estimated lifetime is approximately 199 days, processing an image every 30 seconds. Our study demonstrates that the novel heterogeneous MCU can perform end-to-end CNN inference with an energy consumption of just \SI{4.85}{\milli\joule}, matching the efficiency of the simpler Viola-Jones algorithm and offering power consumption up to 15$\times$ lower than previous methods. Code at: \url{https://github.com/Bomps4/TAFE_Pest_Detection}}
\end{abstract}

\section{Introduction} \label{sec:intro}

The codling moth pest, scientifically known as \textit{Cydia pomonella}, poses a significant threat to global crop production, potentially causing up to 80\% fruit abscission in apple orchards~\cite{JU2021104925,wan2019chromosome}. 
To prevent an excessive utilization of pesticides, it is essential to promptly detect the threat and perform localized treatments only in the affected areas~\cite{agronomy13020324,Irrigation}. 
{Given the size of the areas occupied by the orchard fields, novel automated pest detection systems have become crucial to replace costly and time-consuming human inspection~\cite{suto2022codling}.}

\begin{figure}[t]
\centering
\includegraphics[width=1\linewidth]{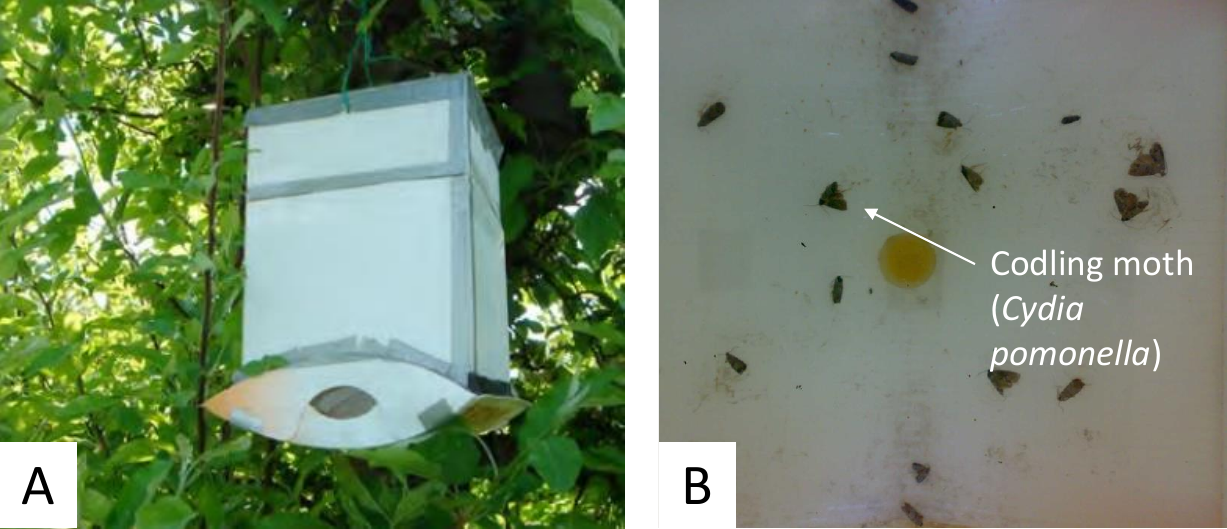}
\caption{A) Trap prototype for codling moths proposed in~\cite{guarnieri2011automatic} and B) an image sample taken from a trap deployed in the field~\cite{guarnieri2011automatic,Brunelli_ours}.}
\label{fig:intro}
\end{figure}

Fig.~\ref{fig:intro}-A shows a trap commonly used for codling moth detection~\cite{guarnieri2011automatic}. 
The prototype contains a sticky pad with a sex pheromone bait to attract the pests.
An internal camera records the trapped insects and sends the images (e.g., Fig.~\ref{fig:intro}-B) to a remote server for analysis.
The presence of the pests is assessed by detecting the codling moths using computer vision algorithms~\cite{survey_methods}. 

Because of the nature of the installation environment, the sensor system can only be powered by batteries or energy harvesters, e.g., solar panels. 
This requirement severely constrains the energy the individual electronic components like the image sensor, the central processing unit, and the communication module can consume. 
In particular, long-range transmissions of images typically require power-hungry interfaces, such as a Global System for Mobile Communications (GSM) subsystem. 
As reported by \textit{López et al.}, the GSM interface consumes 1.5 Watts in the active state, which accounts for 66\% of the total system power~\cite{Lopez_energy}, thus significantly impacting the lifetime of the battery-operated sensor (if we consider one image per hour it would account for $\sim$20 months of operation of the sensor).

Thanks to this \textit{near-sensor processing} approach, only a few bytes of information reporting the pest counts are occasionally sent to the remote station. 
The low data rate enables the usage of ultra-low power protocols such as LoRa~\cite{LoRa}, thus reducing the communication power cost compared to naive \textit{sense-and-transmit} systems~\cite{preti2021developing}. 

On the processing side, Microcontroller units (MCUs) are commonly selected as the core devices of low-power sensor nodes~\cite{Low_power_MCU}. 
These devices include single or multiple CPU cores clocked at up to a few hundred \SI{}{\mega\hertz} and up to a few \SI{}{\mega\byte} of on-chip memory. 
In our recent work~\cite{Ours_old}, we showed an optimized Viola-Jones detector for pest detection on the GAP8 System-on-Chip (SoC), an energy-efficient MCU featuring a compute cluster with eight general-purpose RISC-V cores. 
This method achieves a processing latency 51\% lower than a previous computer vision pipeline running on the same SoC~\cite{Brunelli_ours}, which combined a background subtraction algorithm to localize the pests with a Convolutional Neural Network (CNN) for classifying the detected objects.
{
This paper extends our precedent study~\cite{Ours_old} by analyzing traditional vs. CNN-based image object detection algorithms for on-device pest detection on a State-of-the-Art (SoA) MCU featuring a convolution accelerator. 
This class of processing devices, not yet considered in the context of pest-detection systems, has been recently introduced to speed up onboard analytics tasks by 10-100$\times$ with respect to utilizing only a single general-purpose core (e.g., $\sim$56$\times$ in~\cite{Marsellus} for CNN inference).
\textbf{We selected the GAP9 SoC, an energy-efficient MCU that combines 10 general-purpose cores with a dedicated hardware engine for CNN processing, as an example of this novel class of heterogeneous platforms}. We compare the efficient execution of our optimized Viola-Jones detector with a CNN-based image object detector, namely MobileNetV3-SSDLite (MBNV3-SSD), featuring \SI{3.44} million parameters and a computational load of \SI{584}{\mega\nothing} Multiply-Accumulate (MAC) operations. Finally, the processing unit is coupled with a low-power image sensor and a LoRa transmission module.

In summary, this paper makes the following contributions:
\begin{itemize}
    \item we compare a traditional Viola-Jones detector vs. a CNN-based approach for detecting \textit{Cydia pomonella};
    \item we analyze latency-optimized versions of the two methods when running on the SoA GAP9 SoC;
    \item we evaluate the system-level costs of the pest detection sensor using a duty-cycling power management scheme.
\end{itemize}}

Our study shows that the Viola-Jones detector and MBNV3-SSD reach a similar detection rate of, respectively, \SI{81.7}{\nothing}\% and \SI{83}{\nothing}\% on a first pest dataset. 
For a second set of images with smaller-sized insects, the CNN-based detector can generalize better than Viola-Jones, achieving a detection rate of \SI{72}{\nothing}\%, +\SI{27}{\nothing}\% than the other method. 
On the GAP9 SoC, our Viola-Jones algorithm completes the detection task in \SI{221}{\milli\second} over a 320$\times$240 image, which is 1.47$\times$ faster than the execution time on GAP8, thanks to the higher clock speed (+\SI{37}{\nothing}\%) and the larger memory (+\SI{64}{\kilo\byte} of low-level memory) of the most recent processing unit.
On the other side, the MBNV3-SSD execution takes \SI{147}{\milli\second} when using the convolution accelerator, 9.5$\times$ faster than on the GAP8, which only uses the available general-purpose cores to run CNN workloads.
This speed-up is mainly explained by the higher inference efficiency of GAP9 vs. GAP8, achieving a performance level of \SI{10.9}{MAC/cycle} (7.2$\times$ higher than GAP8). 
Thus, our study indicates that the GAP9 heterogeneous platform can enable on-board processing of accurate CNN-based pest detectors, which was previously shown only on processing units with more than 15$\times$ higher power consumption~\cite{sutHo2021embedded}. 
{ From a system-level perspective, we estimated a battery lifetime of $\sim$199 days if the system is powered by a \SI{1000}{\milli\ampere\hour} battery at \SI{3.7}{\volt} and an image is analyzed every 30 seconds with a duty-cycled power management scheme.}

\section{Related Work} \label{sec:related}

Computer Vision for codling moth detection can be classified into two broad categories: methods running at the edge, i.e., using the processing unit inside the trap, vs. image analysis tasks offloaded to a remote station~\cite{suto2022codling,survey_methods}.
The latter typically adopt CNN-based algorithms to analyze the images transmitted from the wireless camera sensors deployed in-field~\cite{albanese2021automated,vcirjak2023efficientdet,ding2016automatic, sutHo2021embedded, suto2022novel}.
Commercial solutions, like \textit{TrapView}\footnote{\url{https://trapview.com/}} and \textit{iSCOUT}\footnote{\url{https://metos.at/en/iscout/}} offer software packages for server-side analysis of few images per day. 
The system proposed by \textit{Preti et al.}~\cite{preti2021developing} sends images to a central server through the cellular network; the pest detection relies on a particle filter detector, which, based on the morphological data of \textit{Cydia pomonella} provides regions of interests in the image, later classified by a CNN. All have been implemented with the \textit{ImageJ} software package for image analysis~\cite{Schneider2012}.
In~\cite{suarez2021pest}, the authors proposed a two-stage computer vision pipeline to process the images received from the traps. A canny-edge detector is followed by a CNN for the classification task, resulting in an overall accuracy of 94.8\%.
In contrast to this class of methods, our solution follows a \textit{near-sensor processing} paradigm, where image analysis is performed directly in the trap to reduce the transmission bandwidth requirements and the associated communication energy and financial cost (if a metered network is used).

Among the solutions proposed for on-board processing, \textit{Suto}~\cite{suto2022novel} combined a Raspberry PI Zero W board for image acquisition and analysis with a LoRa module to communicate the number of detected objects. 
The processing task consists of a Selective Search (SS) algorithm and a CNN classifier, MobileNetV2, leading to a total compute cycle of 195 seconds. 
Following the pipeline proposed by \cite{ding2016automatic}, the work by \textit{Albanese et al.}~\cite{albanese2021automated} leverages a CNN classifier (LeNet is the more efficient setting among the ones proposed) to classify the image patches selected by a moving window. 
The system revolves around a Raspberry PI 3 board, augmented with a dedicated accelerator board, the Intel Neural Compute Stick (NCS), to speed up the CNN inference workloads. 
The method reached an accuracy of up to 96\% at a total power cost of \SI{5}{\watt} (\SI{3}{\watt} for the main board and \SI{2}{\watt} for the NCS), measuring a lifetime of up to 3 months if processing two images per day and sending the count of detected objects to the server using LoRa. 
A similar smart trap design was proposed in~\cite{vcirjak2023efficientdet}: a Raspberry PI 4 (RPI4) coupled with a \SI{12.5}{\mega pixel} camera runs an EfficientDet CNN-based object detector and then sends the number of detected pests. 
This solution achieved an accuracy of 99.3\% on their test data; the system turns on once a day, reaching up to 5W of power consumption. 
Despite their effectiveness, these methods rely on high-end embedded processors featuring a power consumption of several Watts, more than 10$\times$ higher than our target. 

In the context of ultra-low-power systems, \textit{Schrader et al.}~\cite{schrader2022open} proposed to use an ESP32 device with a power envelope of \SI{300}{\milli\watt} for image capturing.
The final system, however, is solely employed to send the taken pictures to a central station for processing. 
The work in~\cite{sutHo2021embedded} used the OpenMV~\cite{abdelkader2017openmv} board with a high-performance STM32H7 MCU, clocked at \SI{480}{\mega\hertz}, and a VGA OV7725 image sensor. 
The MCU runs an SS kernel to localize the insects inside the image and a MobileNetV2 for the classification, reaching 82\% accuracy. 
The system stays in active mode for $\sim$1 minute a day, consuming more than \SI{500}{\milli\watt} to perform image capture, object detection, and sending the pest counts through GSM.

To reduce the power consumption, the solution of \textit{Brunelli et al.}~\cite{Brunelli_ours} revolves around the ultra-low power multi-core GAP8 MCU, which executes a background subtraction algorithm based on a Gaussian Mixture Model (GMM) to propose regions of interest that are later classified by a CNN algorithm, the latter classification algorithm achieves 93\% accuracy. 
This system uses the Long Range Wide Area Network (LoRaWAN) protocol to send the images when detection is performed, with an energy cost of up to \SI{52}{\joule} per image.
Unlike these works that rely on two-step computer vision pipelines, we recently proposed a simplified single-stage object detector based on the Viola-Jones algorithm that directly returns the pests' locations in the image~\cite{Ours_old}. 
On GAP8, this method takes only \SI{400}{\milli\second} to run on the 8-core cluster, $\sim$2$\times$ faster than~\cite{Brunelli_ours}. 

This paper differentiates from these previous studies by comparing our traditional Viola-Jones algorithm and a CNN-based object detector, also adopted by high-end systems~\cite{albanese2021automated,sutHo2021embedded}, using -- for the first time in the context of pest detection systems -- a novel heterogeneous MCU with a specialized CNN accelerator.
Thanks to the increased processing efficiency compared to the precedent MCU generation, the more complex CNN algorithm shows a similar energy consumption than the traditional detector  (\SI{4.85}{\milli\joule} vs. \SI{4.61}{\milli\joule}), bridging the gap with more power-hungry systems.
\section{Background} \label{sec:background}

\subsection{Viola-Jones} \label{sec:Viola_jones}

The Viola-Jones algorithm~\cite{viola2001rapid} scans an image with a moving window and checks for the presence of the target objects. 
The algorithm extracts the hand-crafted Haar-like features for every image window by applying a set of rectangular filters and then runs a multi-stage cascade classifier.
The classifier consists of many simple and computationally inexpensive tests, denoted as \textit{weak classifiers}, which operate on the Haar-like features.
Every stage of the cascade classifier includes multiple weak classifiers contributing to a final stage's score. 
If this score is below a threshold, the detection response on the current window is negative, and no detection is performed. 
On the other side, a detection is made if all the cascade stages are successfully passed. 
The set of filters of every stage and the threshold values are determined by a training process fed by positive and negative samples. 
Instead of operating on the image space, the original paper proposed an intermediate representation, denoted as the \textit{Integral Image} $II$, to compute the Haar-like features efficiently. 
{
This integral image stores at every location  $(x,y)$ the sum of the pixels of the input image, i.e., the total brightness level, in the rectangular area delimited by $(0,0)$ and $(x,y)$. For example, in figure~\ref{fig:Integral_image}, the pixel A in the integral Image, $II[A]$, is the sum of the pixels in the area A of the input image. By precomputing the integral image, we can easily calculate the integral of any arbitrary area in the input image, which is the founding operation for the Haar-like features.

\begin{figure}
    \centering
    \includegraphics[width=\columnwidth]{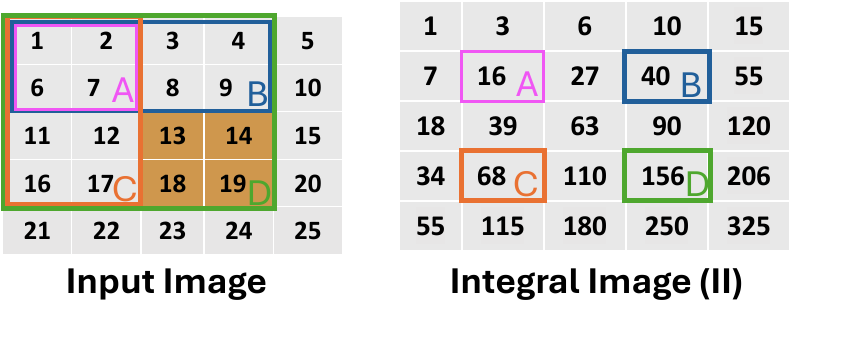}
    \caption{Example of the input image (left) and the corresponding integral image (on the right). The values A-D of the integral image correspond to the integral of the areas A-D in the input image (rectangles with matching letters and colors). By precomputing the integral image, the integral value of the brown area can be computed as II[A] -II[B]-II[C] +II[D].}
    \label{fig:Integral_image}
\end{figure}

For example, the integral of the brown area in figure~\ref{fig:Integral_image}, is simply computed as $II[D] + II[A] - II[B] -II[C] =$~$156 +16 - 40 - 68 = 32$: the operation requires only 4 read accesses in the memory independently of the size of the integral area.
}

\subsection{MobileNetV3-SSDLite} \label{sec:mbv3}
{Following the analysis performed by~\cite{Butera}, we selected the MobilenetV3-SSDLite model for our pest detection task as it represents the best compromise between accuracy and network complexity. Other models with higher accuracy scores cannot fit the memory resources (e.g., Retina-Net~\cite{RetinaNet}) of our platforms or do not feature predictable execution times (e.g., Faster R-CNN~\cite{FasterRCNN}), which is essential for real-time embedded devices.}
MobileNetV3-SSDLite is a CNN-based object detector that is composed of a feature extractor, also denoted as the \textit{backbone}, and a regression (or classification) head~\cite{SSD}.
More in detail, this model uses MobileNetV3~\cite{Mobilev3} as the feature extractor and the SSDLite as the regression head~\cite{SSD}. 
Both the sub-networks consist of multiple convolution layers. 
Each layer applies a convolution filter over an input feature map, typically denoted as the \textit{activation} tensor. 
The activation tensor output is then passed to the next layer.
Hence, the final network result is obtained by propagating data through all the layers, also called the \textit{inference} task. 

The MobileNetV3 and the SSDLite blocks rely on depthwise-separable convolution as the fundamental operator. 
Concerning standard 2D convolutional filters, a depthwise-separable convolution is composed of a depthwise and pointwise convolution, reducing the number of filter parameters and operations by 86.8\% in the case of a $3\times3$ convolution with 16 input and output channels.
More in detail, the MobileNetV3 architecture includes Inverse Residual Blocks, which expands the number of input features with a pointwise convolution. 
The produced feature maps are then passed to a depthwise convolution and finally compressed back with another pointwise layer. 
A residual connection is added when the input feature size matches the size of the output tensor.
On the other side, SSDLite comprises four depthwise-separable convolution blocks, each attending a diverse object size.
This regression head takes the features produced by the backbone and produces the bounding boxes and classes of the object inside the image.
\section{Methodology} \label{sec:methodology}

\subsection{On-device Processing Requirements}
To determine the throughput requirement for on-device pest detection, we examine the operational role of a smart trap deployed in the field. Smart traps are instrumental in relaying data on pest activity, which is crucial for implementing timely and effective pest control measures. Based on this, we define two application scenarios dictating the latency requirement for the processing task:
\begin{itemize}
\item \textbf{Low-energy scenario.} In this scenario, we perform detections at the same frequency as pheromone sprays, one every 15 minutes~\cite{Emission_aerosol}, ensuring our platform's responsiveness while minimizing energy consumption. Thanks to this, we can provide information on pest activity, which can be used to optimize pheromone spraying, releasing higher concentrations only when pest activity is detected and reducing unnecessary pheromone waste, as proposed in ~\cite{pheromone}.
\item \textbf{High-frequency scenario.} In this scenario, we increase the detection frequency to one every 30 seconds to address the limitations of current pest management methods, which rely on daily assessments of pest presence and temperature~\cite{simulation_population,Cidia_lifetime}. This higher detection rate fills a data gap, allowing for more accurate analysis by correlating pest activity with environmental factors like airspeed and humidity. Since the presence of even a single moth can trigger pest disruption methods~\cite{codling_emergence}, the capability to detect each new moth entering the trap is necessary to improve our understanding of pest behavior and improve current pest management techniques.
\end{itemize}
In both scenarios, the system must immediately transmit the pest count after detection to activate the pheromone dispensers. From a system-level viewpoint, we target a battery-powered node that can operate for the whole activity period of the Cydia pomonella, around 4 months~\cite{Cidia_lifetime}. Due to its cost-effectiveness and low cost (less than 10 dollars), we have selected a 1000 mAh battery as a system requirement because this type of battery is commonly used in low-power sensor nodes~\cite{Node_1_battery,Node_2_battery} and IoT sensor nodes \cite{Node_iot_battery}.

\subsection{Heterogeneous Processing Platform}

\begin{figure}[t]
\centering
\includegraphics[width=\columnwidth]{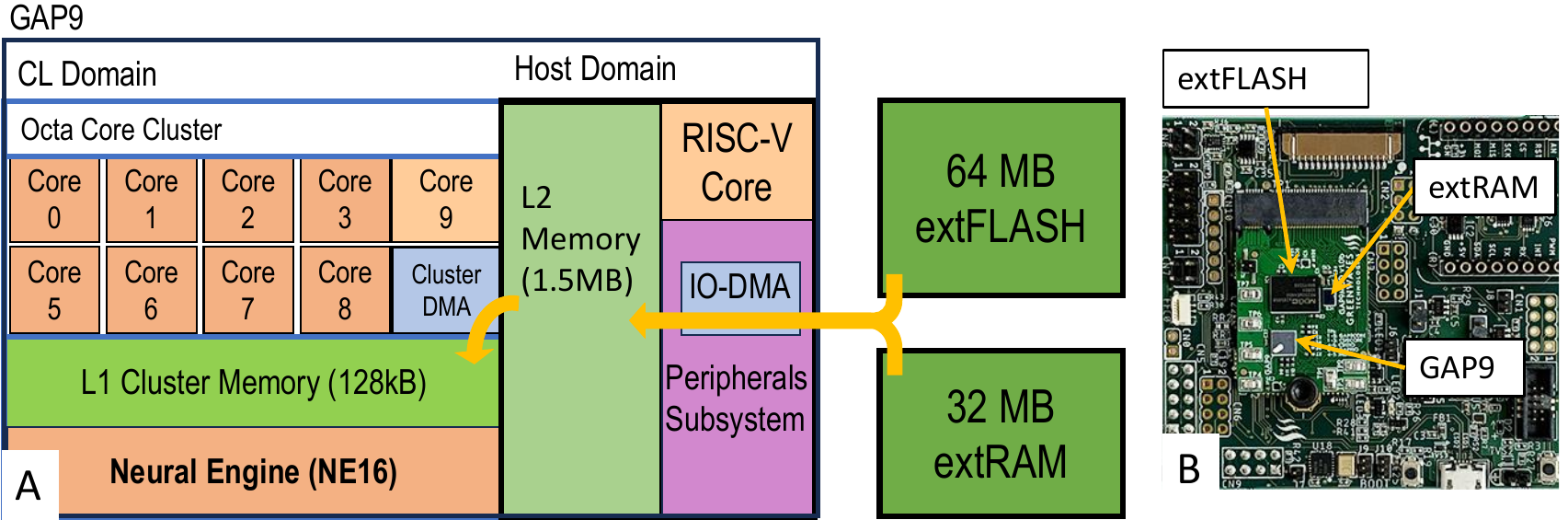}
\caption{A) Block diagram of the GAP9 SoC and the interfaces with the external memories. The yellow arrows indicate the data copies between the L1 and L2 memories and between the off-chip memories and the L2 memory operated by, respectively, the Cluster DMA and the IO-DMA. B) GAP9 carrier board.}
\label{fig:decoding}
\end{figure}

\begin{table}[t]
    \centering
    \caption{Comparison between the GAP8 and the GAP9 architectures.}
    \label{tab:comparisonArchitectures}
    \resizebox{1\linewidth}{!}{
    \begin{tabular}{lcc}
    \toprule
    \textbf{}&\textbf{GAP8}&\textbf{GAP9}\\
    \midrule
    {RISC-V Cores \textit{Host} + \textit{Cluster}}& 1+8 & 1+9\\
    {On-chip L1 memory}&\SI{64}{\kilo\byte}&\SI{128}{\kilo\byte}\\
    {On-chip L2 memory}&\SI{512}{\kilo\byte}&\SI{1.5}{\mega\byte}\\
    {CNN Accelerator}&No&NE16\\
    {Floating point hardware support}&No&Yes\\
    {VDD}&\SI{1.2}{\volt}&\SI{0.65}{\volt}\\
    {Max Clock Frequency \textit{Host}}&\SI{250}{\mega\hertz}&\SI{240}{\mega\hertz}\\
    {Max Clock Frequency \textit{Cluster}}&\SI{175}{\mega\hertz}&\SI{240}{\mega\hertz}\\
    Fabrication Technology& \SI{65}{\nano\meter} & \SI{22}{\nano\meter}\\
    Peak Efficiency& \SI{4.24}{\milli\watt/\giga OP} & \SI{0.33}{\milli\watt/\giga OP}\\
    \bottomrule
    \end{tabular}}
\end{table}

We use the GAP9 SoC by Greenwaves Technologies as the main processing unit for on-board pest detection. 
The micro-architecture of GAP9, which is schematized in Fig.~\ref{fig:decoding}A, is composed of a \textit{Host} domain and a programmable accelerator sub-system, denoted as \textit{Cluster} (CL). 
The \textit{Host} domain includes a RISC-V programmable core, a \SI{1.5}{\mega\byte} L2 memory, and a broad set of peripherals. 
Conversely, the cluster combines 9 RISC-V cores with a dedicated convolution accelerator, called NE16, to accelerate the execution of compute-intensive linear algebra kernels, e.g., CNN's convolution operations.

The RISC-V cores feature vectorial MAC instructions that execute a dot-product between two 4$\times$8-bit vectors and accumulate the results in a single clock cycle. 
The NE16 comprises instead 9$\times$9$\times$16 8$\times$1bit MAC units, reaching a peak compute efficiency of 150 8-bit MAC operations per clock cycle, up to $\sim$5$\times$ faster than using the cluster cores. 
The cluster domain also features a low-level L1 scratch memory of \SI{128}{\kilo\byte} that can be accessed in a single-clock cycle from the cores and NE16. 
These engines can also access data stored in the larger L2 memory but with a $\sim$10$\times$ lower bandwidth than accessing the L1 memory.
Alternatively, a cluster DMA engine can be programmed to copy data between the L1 and the L2 memories in the background of the core operations. 

At the board level (Fig.~\ref{fig:decoding}B), the chip is interfaced with an external \SI{32}{\mega\byte} RAM and a \SI{64}{\mega\byte} FLASH memory via octaSPI. 
An IO-DMA within the peripheral sub-system of the GAP9's \textit{Host} domain is responsible for copying data between the external and on-chip L2 memory with a maximum data rate of 1 byte per clock cycle. 
1D or 2D memory transfers can be programmed via software; the software call of a 2D data copy enqueues multiple 1D copies. 
In GAP9, the \textit{Host} and the cluster reside on separate voltage and frequency domains, enabling fine-grained dynamic voltage and frequency scaling. 
To gain maximum energy efficiency, we configure a voltage level of \SI{0.65}{\volt} and a clock frequency of \SI{240}{\mega\hertz} for both domains. 

Tab.~\ref{tab:comparisonArchitectures} summarizes the main characteristics of the GAP9 and highlights the main differences concerning the precedent MCU generation, the GAP8 SoC, which was used for image-based pest detection by some recent works~\cite{Brunelli_ours,Ours_old}.
Notably, GAP9 features a higher on-chip memory budget and a specialized hardware engine to speed up CNN inference workloads absent on GAP8.

\subsection{Accelerating Pest Detection on GAP9}
This section describes the optimized deployment strategies of the pest detection algorithms on GAP9. 
Both the Viola-Jones and the MBNV3-SSD CNN detectors are trained on a custom dataset composed of 158 images (320$\times$240) with a total of 1275 target insects manually labeled.
{  All the images were taken using an RGB sensor with a resolution of 2048$\times$1536 pixels under daylight conditions (from 8 am to 7 pm) without employing any external illumination. }Because the collected images contain 27 non-target objects ($<$1.6 \% of the total amount of pests), of which only 3 in the test split, our detection algorithms are designed to identify only objects belonging to the \textit{Cydia pomonella} category.

\subsubsection{Viola-Jones}
A 15-stage cascade classifier is trained using a variant of the AdaBoost algorithm~\cite{viola2001rapid}.
The training set is composed of the 20$\times$20 patches of \textit{Cydia pomonella} cropped from the available images and a set of negative crops from a different image dataset available online\footnote{\url{https://pythonprogramming.net/static/images/opencv/negative-background-images.zip}}.
To account for the diverse sizes of the target objects, we run the Viola-Jones algorithm over five scales of the original images.
At every iteration, the image resolution is scaled down by a factor of 1.1 in horizontal and vertical dimensions.
Detections larger than 30$\times$30 pixels are discarded.

To run the Viola-Jones detector on the GAP9 MCU efficiently, we use the parallel software code presented in our recent work~\cite{Ours_old}.
In this implementation, the input image is stored in the L2 memory, occupying \SI{76}{\kilo\byte} (320$\times$240 \texttt{INT8}).
Unfortunately, the L1 memory is not large enough to fit the input data and the integral image, which demands a total of \SI{307}{\kilo\byte} (320$\times$240$\times$4 byte/elements -- \texttt{INT32} values).
To solve this problem, we split the original image (and the scaled versions) into tiles of size 100$\times$240 pixels. 
Every tile is copied from the L2 to the L1 memory at runtime and then processed. 
To account for potential detections between adjacent tiles, we set an overlap of 20 pixels in the tiling logic for both the horizontal and vertical directions.

The processing kernel takes an input data tile to compute a tiled version of the integral image, which features a size of \SI{96}{\kilo\byte} and is stored in the L1 memory. 
The CL cores then scan (in parallel) different regions of the image tile with a 20$\times$20 sliding window.
Every core invokes the cascade classifier to test if the current window includes the target object. 
When a stage of the classifier returns a negative response, the evaluation continues to the next window of the tile.
The parallelization is operated over eight cluster cores; the ninth core works as the task dispatcher. 
At the end of the analysis, the algorithm outputs a list of detections described by their bounding box coordinates.

\subsubsection{Convolutional Neural Network}
The MobileNetV3-SSDLite object detection network features \SI{3.44}{\mega\nothing} parameters and accounts for \SI{584}{\mega\nothing} MAC operations to process a 320$\times$240 image. 
Similarly to Viola-Jones, this detector outputs a set of bounding boxes, each associated with a predicted class, i.e., \textit{Cydia Pomonella} vs. \textit{background}.
The model is trained for five epochs using our custom dataset with an 80-20\% split between training and validation, resulting in 127 and 31 images, respectively.
Stochastic Gradient Descent (SGD) is the optimizer with a momentum of 0.9, a weight decay of $5*10^{-4}$, and a learning rate of 0.005.
The weights of the MobileNetV3 backbone are initialized from a model pre-trained on the COCO dataset~\cite{lin2014microsoft}; the parameters of the SSDLite heads are instead learned from scratch. 
To deploy the CNN detector on GAP9, we use the \textit{GAP}flow toolset\footnote{https://greenwaves-technologies.com/tools-and-software/}.
This software package takes the CNN model description (\texttt{ONNX} format) and generates the inference \texttt{C} code.
To use the NE16 accelerator, the model is first quantized to 8-bit using the post-training quantization routine of the \textit{GAP}flow, which takes a few samples from the training set (2 in our case) to estimate the quantization range, i.e., the \textit{calibration} process. 
The 8-bit quantized parameters are then stored in the external FLASH memory. 

At runtime, the inference program follows a layer-by-layer processing schedule. 
For this task, we allocate one memory buffer in the L2 memory to store the layer operands and a buffer in the L1 memory, serving to store temporary results. 
The sizes of these buffers are, respectively, \SI{1.2}{\mega\byte} and \SI{115.6}{\kilo\byte}. 
Depending on the layer requirement, the program can transfer the weight parameters from the external FLASH memory to the on-chip L2 memory before the layer execution. 
Conversely, a layer's input and output activation tensors may be allocated in the on-chip L2 fast memory or the external RAM.

Every layer function copies data (weights and activations) from the L2 (or external) memory to the L1 buffer. It dispatches the processing job, e.g., a convolution operation, on the NE16 engine or the general-purpose cores.
The latter option is selected if no convolution accelerator is available, like in the GAP8 SoC, or if the accelerator does not support that layer.
We also highlight the impact of the L1 and L2 buffer sizes on the inference throughput. 
Generally, a larger L2 memory buffer favors allocating more tensors on the on-chip memory, leading to a faster read and write time than accessing data to/from off-chip memories. 
At the same time, an increased L1 buffer reduces the number of copies from the L2 memory, lowering the associated latency overheads. 

\begin{figure}[t]
\centering
\includegraphics[width=1\linewidth]{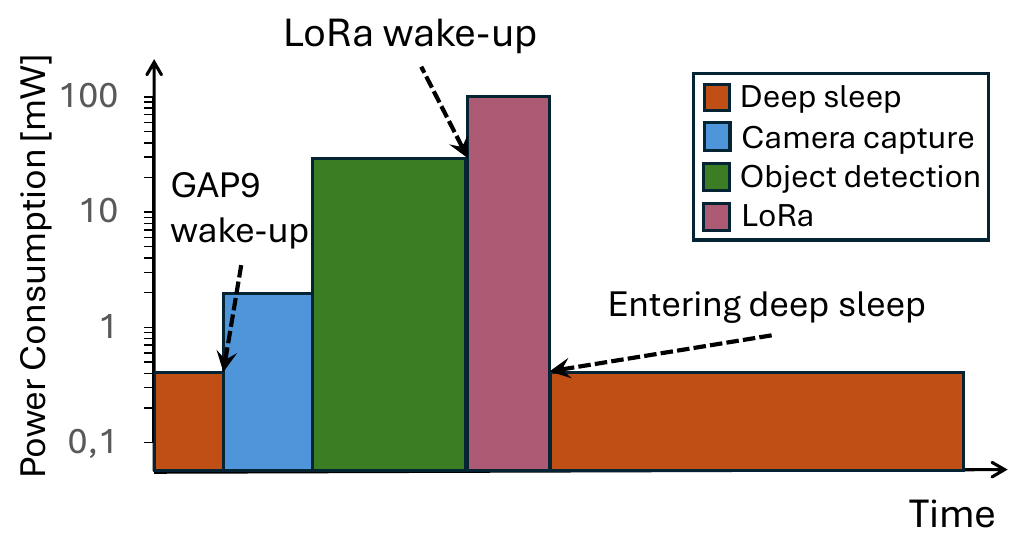}
\caption{System-level power management strategy. The system periodically wakes up from deep sleep mode to capture and process an image. Then, the output is sent via LoRa before going back to deep sleep.}
\label{fig:order}
\end{figure}

\subsection{System-level Power Management} \label{sec:system}
{The final pest detection system includes the GAP9 processing unit, a Himax HM01B0 ultra-low-power camera sensor, which captures 320$\times$240 pixels 8-bit grayscale images, and a LoRa module, i.e., the HTCC-AB01 board by CubeCell, connected via UART with the MCU.}
The latter relies on the AM01 module, which incorporates a low-power cortex M0 MCU with \SI{16}{\kilo\byte} of RAM running at \SI{48}{\mega\hertz} and the SX1262 transceiver implementing the physical frequency modulation of the LoRa standard.

Fig.~\ref{fig:order} shows the used power management scheme to minimize the system energy consumption. 
Following a duty-cycling scheme, the MCU wakes from deep sleep using the internal real-time clock (RTC).
In the active state, the system captures an image and runs the pest detection algorithm. 
Lastly, the GAP9 MCU activates the HTCC-AB01 board module to transmit the count of detected insects using the LoRa protocol at an energy cost of \SI{1}{\milli\joule} per byte sent. 
Once the transmission is completed, the GAP9 receives an acknowledgment signal from the HTCC-AM01 module and returns to the deep sleep mode. 
We use this same execution scheme for both our low-energy and high-frequency scenarios.
\section{Experimental Results}\label{sec:results}

\begin{table}[t]
\caption{Detection accuracy on the \textit{near-ds} and \textit{far-ds} datasets.}
\label{tab:accuracy}
    \centering
    \begin{tabular}{cccc}
    \toprule
        Method & Quantization & \textit{near-ds} & \textit{far-ds} \\ \midrule
        Viola-Jones & \texttt{int8} & \SI{81.7}{\nothing}\% & \SI{45.0}{\nothing}\% \\ \hline
        MBNV3-SSD & \texttt{float} & \SI{88.0}{\nothing}\% & \SI{84.0}{\nothing}\% \\ 
        MBNV3-SSD & \texttt{int8} & \SI{83.0}{\nothing}\% & \SI{72.0}{\nothing}\% \\ \bottomrule
    \end{tabular}
\end{table}

\subsection{Detection Accuracy}
We tested the effectiveness of the Viola-Jones and the MBNV3-SSD detectors on two test sets, namely \textit{near-ds}, which includes 11 samples with 196 target objects, and \textit{far-ds}, composed of 7 images for a total of 158 objects.
The two datasets differ by the distance between the sticky pad and the camera sensor.
All the images were taken using an RGB sensor with a resolution of 2048$\times$1536 pixels. 
{For a fair evaluation of the characteristics of the low-power sensor of our system, we resize the images of both datasets to 320$\times$240 pixels and apply a greyscale transformation with additive Gaussian noise to mimic the image quality of the Himax camera sensor.}

\begin{figure}[t]
\centering
\includegraphics[width=1\linewidth]{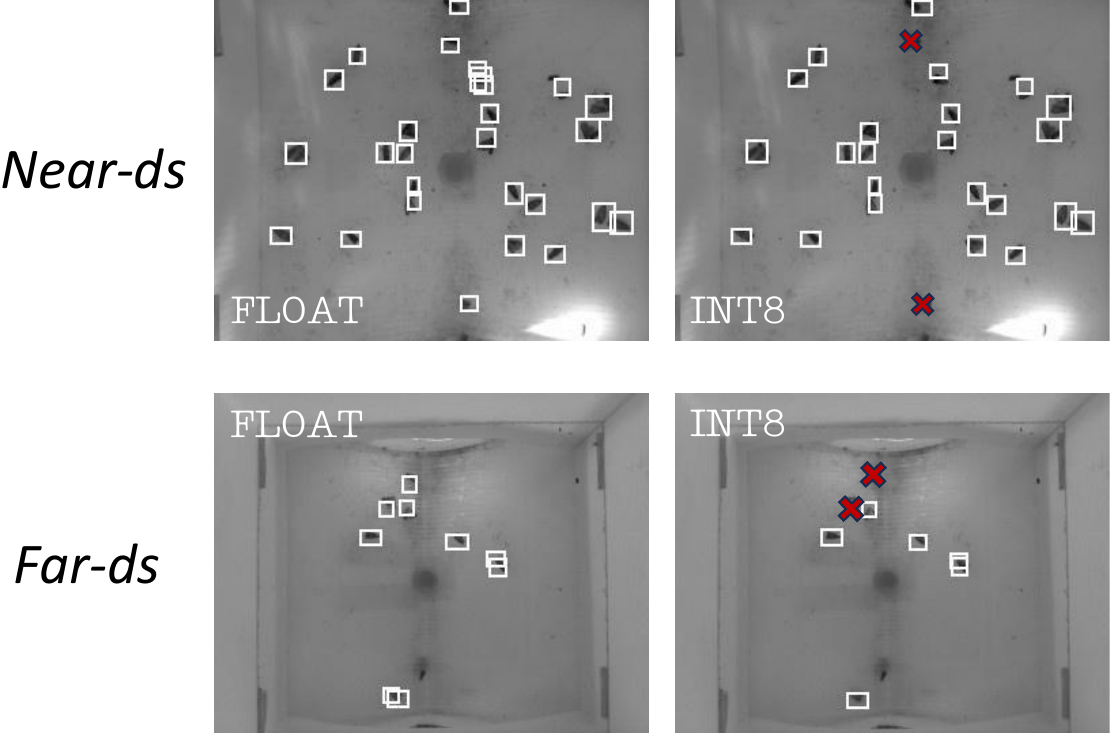}
\caption{CNN-based pest detection on the image samples from the \textit{near-ds} (upper row) and \textit{far-ds} (bottom row) testsets. Outputs from the baseline \texttt{float} model (left) are visually compared with the detections of the \texttt{int8} quantized model. The red crosses mark the miss-detections.}
\label{fig:det_comp}
\end{figure}

Tab.~\ref{tab:accuracy} reports the detection accuracies scored by Viola-Jones and MBNV3-SSD when in full-precision (\texttt{float}) or quantized to 8-bit (\texttt{int8}). 
Our metric assesses a correct detection if the predicted bounding box overlaps a manual annotation (the IoU score used by the COCO metrics~\cite{cocodataset} is set to 0.01).
The Viola-Jones algorithm achieves a detection accuracy of 81.7\% on the \textit{near-ds} but only scores 45\% on the \textit{far-ds} dataset.
We motivate this gap with the poor generalization capacity of the Viola-Jones algorithm: the training set shows a higher correlation with the \textit{near-ds} than the \textit{far-ds} set.
On the other side, the \texttt{float} baseline of the MBNV3-SSD detector reaches a superior detection accuracy compared to Viola-Jones in both datasets: 88.0\% and 84.0\% for, respectively, the \textit{near-ds} and \textit{far-ds}.
The post-training quantization process introduces an accuracy drop, reaching \SI{-12.0}{\nothing}\% for the \textit{far-ds} case. 
On the other dataset, the degradation in performance is limited to \SI{-5}{\nothing}\%, possibly also explained by the reduced difference between the training and test domains.
As an illustrative example, Fig.~\ref{fig:det_comp} shows the outputs of the \texttt{float} vs. \texttt{int8} models on reference samples from the two sets, highlighting the miss-detections of the quantized model.

\subsection{Latency Analysis}
Tab.~\ref{tab:memory_lat} reports the memory usage and the latency measured when executing the Viola-Jones and MBNV3-SSD on the GAP9 SoC with the NE16 accelerator. 
For comparison purposes, we also report the measurements on the GAP8 MCU, where the CNN inference task runs on the eight general-purpose cores of the cluster (indicated by $CL$ in the table).
Both chips are configured in their best energy-efficient point: the clock frequency of GAP9 is set to \SI{240}{\mega\hertz} with an operating voltage of \SI{0.65}{\volt} while GAP8 operates at \SI{175}{\mega\hertz} at \SI{1.2}{\volt}. 

The Viola-Jones algorithm requires \SI{384}{\kilo\byte} of L2 memory for the input and the integral image.
The L1 memory stores an image tile (the tiles size is 100$\times$240 on GAP9 vs. 100$\times$120 on GAP8), with a maximum occupation of \SI{99.6}{\kilo\byte} and \SI{51.6}{\kilo\byte}, for respectively GAP8 and GAP9. 
The parameters of the cascade classifier, also preserved in the low-level memory, occupy only \SI{3.56}{\kilo\byte}. 
No external memory is required. 
On the other hand, MBNV3-SSD uses the off-chip FLASH memory to permanently store the network parameters for a total of \SI{3.44}{\mega\byte}.
As described in Sec.~\ref{sec:mbv3}, we allocate an L2 buffer for (part of) the parameters, the activation tensors, and an L1 array acting as the working memory.
Additionally, an array of \SI{1.6}{\mega\byte} is allocated in the external RAM for the input and output activation tensors not fitting into the L2 memory. 

\begin{figure}[t]
\centering
\includegraphics[width=1\linewidth]{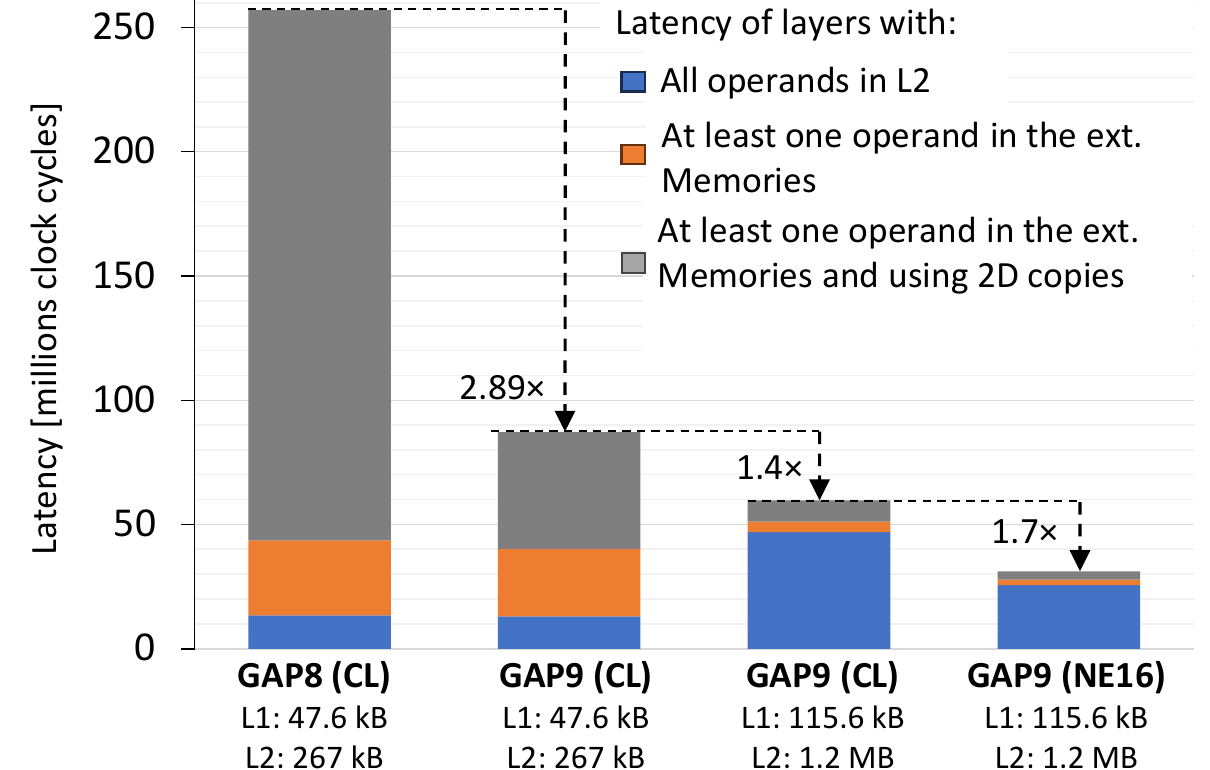}%
\caption{Latency analisys of the MobileNetV3-SSDLite execution on GAP8 and GAP9 under varying L1 and L2 memory budgets. }
\label{fig:ciop_sub}
\end{figure}

Tab.~\ref{tab:memory_lat} also reports the number of clock cycles spent for the algorithm executions. 
For the CNN, we also detail the ratio between MAC and clock cycle (MAC/cycle in the table) as an efficiency indicator. 
Overall, the execution is 1.43$\times$ and 9.5$\times$ faster on GAP9 than executing, respectively, Viola-Jones and MBNV3-SSD on GAP8.
On the latter device, our parallel Viola-Jones software outperforms the CNN execution by 4.43$\times$. 
Thanks to the CNN accelerator, GAP9 achieves a processing time of \SI{147}{\milli\second} for the MBNV3-SSD execution, 1.5$\times$ lower than the Viola-Jones latency. 
Fig.~\ref{fig:ciop_sub} gives insights into the CNN inference latency measurements on GAP8 and GAP9 when varying the sizes of the L2 and L1  buffers and when using the general-purpose cores or the NE16 accelerator as the principal compute engine (indicated, respectively, with CL and NE16 in the figure). 
In the plot, we break down the processing time with respect to the memory locations of the operands of the different layers. 
More in detail, we distinguish between the total latency of the layers whose inputs and outputs are in the L2 memory (indicated as L2 on the plot, in blue) and the layers that fetch input data from the external memory, highlighting the cases that use 1D or 2D memory transfers, respectively in orange and grey. 

When comparing the GAP8 vs. GAP9 execution in the CL setting under the same memory budget (\SI{46.7}{\kilo\byte} of L1 and \SI{267}{\kilo\byte} of L2), the higher clock cycle count on GAP8 is due to not-optimized 2D memory transfers from the external memory (4.6$\times$ slower read time).
When increasing the memory sizes of the L2 and L1 buffers on GAP9, a 1.4$\times$ speed-up is achieved mainly thanks to the higher bandwidth of the on-chip vs. the external memory. 
In this case, the workload of the layers whose inputs are in the L2 memory dominates the total number of operations (93\% of the total) and the total execution time (79\% of the clock cycles). 
These layers' MAC/cycle ratio also increases by +10\% on average concerning the low-memory configuration due to the larger tensors stored in the  L1 memory buffer. 
Lastly, the NE16 accelerator further boosts the system efficiency by 1.7$\times$. 
The speed-up is lower than the theoretical peak because of two main reasons. 
The non-linear \textit{Hsigmoid} function featured by MobileNetV3 is not natively supported by the accelerator. Hence, it must be executed on the general-purpose cores. 
Second, the depthwise convolutions underuse by 1/16 of the NE16 compute engine.

\begin{table}[t]
\caption{Memory Footprint and latency. }
\label{tab:memory_lat}
\resizebox{\linewidth}{!}{
\begin{tabular}{c| c | cccc| cc}
\toprule
Method & Platform & \begin{tabular}[c]{@{}c@{}} L1 \\[0pt] [\SI{}{\kilo\byte}]  \end{tabular}&  \begin{tabular}[c]{@{}c@{}} L2 \\[0pt] [\SI{}{\kilo\byte}] \end{tabular} & \begin{tabular}[c]{@{}c@{}} extRAM \\[0pt] [\SI{}{\mega\byte}] \end{tabular} & \begin{tabular}[c]{@{}c@{}} extFLASH \\[0pt] [\SI{}{\mega\byte}] \end{tabular} & \begin{tabular}[c]{@{}c@{}} Latency \\[0pt] [\SI{}{\mega\nothing} cycles ] \end{tabular} &\begin{tabular}[c]{@{}c@{}} MAC/ \\ cycle \end{tabular}  \\ \midrule
\multirow{2}{*}{\textbf{Viola-Jones}} & GAP8 (CL)                 & 51.6                 & 384                  & --    & --                     & 55.0   & --            \\
                                          & GAP9 (CL)                 & 99.6                & 384                  & --   & --                      & 52.3  & --              \\ \hline
\multirow{2}{*}{\textbf{MBNV3-SSD }}  & GAP8 (CL)           & 47.6                 & 267                  & 1.6  &3.44                     & 255.3        & 1.5       \\ 
                               & GAP9 (NE16)           & 115.6                & 1200                 & 1.3 & 3.44                  & 35.3 & 10.9                \\
                      \bottomrule
   
\end{tabular}%
}
\end{table}%
\subsection{Energy Analysis}
\begin{figure}[t]
\centering
\includegraphics[width=1\linewidth]{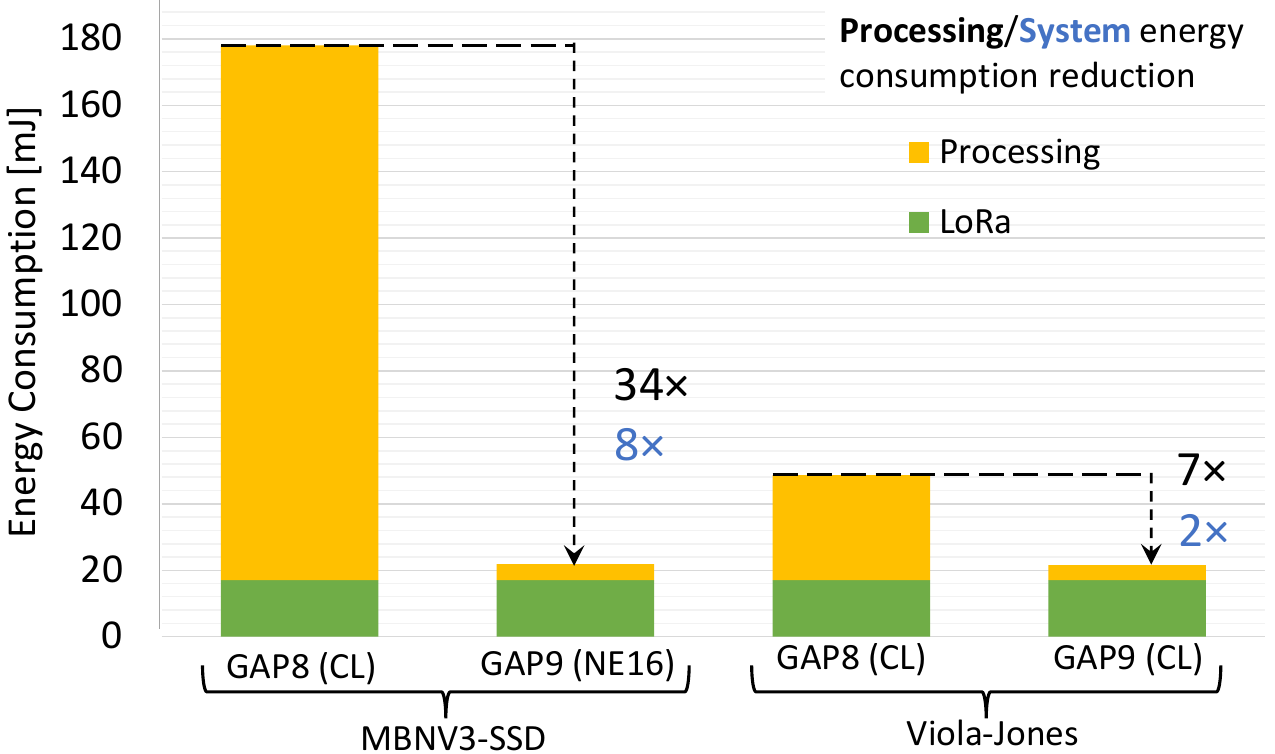}
\caption{Energy consumption of the system during an active cycle composed of camera acquisition, processing, and transmission of the result.}
\label{fig:energy}
\end{figure}

In Fig.~\ref{fig:energy}, we estimate the energy consumption of our system during an active cycle that includes the camera capture, the object detection, and the LoRa wireless transmission. 
In the plot, we break down the energy costs for the image sensor, the MCU, namely GAP8 or GAP9 running Viola-Jones or the MBNV3-SSD, and the LoRA module described in Sec.~\ref{sec:system}. 
First, the energy consumption of the camera sensor is negligible with respect to the other system costs. 
On GAP8, the CNN compute cost is $\sim$10$\times$ higher than the transmission cost (\SI{161}{\milli\joule} vs \SI{17}{\milli\joule}). 
We account for the radio's cost to transmit 4 bytes for the pest count and 13 bytes of the LoRaWAN protocol overhead. 
With Viola-Jones, the energy consumption gap decreases, but the processing (\SI{31.8}{\milli\joule}) still presents the largest energy cost. 
On the other hand, the energy consumption of the GAP9 SoC, which we measured on an evaluation board comprising the external memories, is 7$\times$ and 34$\times$ lower than the energy consumption of the GAP8 MCU for the execution of, respectively, Viola-Jones and MBNV3-SSD. 
The 59\% of this improvement is explained by the micro-architectural innovations of GAP9 vs. GAP8, i.e., more memory and the NE16 accelerator, reducing the clock cycle count. 

The rest can be instead attributed to the lower power consumption of GAP9, reaching \SI{20.5}{\milli\watt} and \SI{33}{\milli\watt} for Viola-Jones and the CNN processing, the latter including also the energy consumption of the external memories. 
On GAP8, we measured a power consumption of \SI{79}{\milli\watt} instead. 
{ From a system-level viewpoint, the energy cost of the GAP9 processing unit is approximately $\sim 3.5 \times$ lower than the transmission cost, thus enabling energy-efficient near-sensor pest detection with a minimal impact on the overall energy cost. This solution presents a consumption higher of only 0.24 mJ with respect to the lightweight Viola-Jones algorithm while providing a higher accuracy (+1.3\% and +27\% respectively for the \textit{near-ds} and \textit{far-ds})}

\subsection{Comparison vs. State-of-the-Art}
{ We evaluate the energy consumption of our method in a complete system composed of a camera, the MCU with its external memories, and a LoRa transmitter as described in section~\ref{sec:system}. 
Tab.~\ref{tab:comparison} compares our solution with other SoA camera-based systems for codling moth detection that use MCUs for onboard processing.
{The table outlines the computer vision algorithms employed and their processing and transmission costs during the active state. It also details the total daily energy consumption, which accounts for the energy spent in active and deep sleep states, and the energy needed to load the DNN’s weights from the external to the on-device memory.} 
We adopt a transmission policy for the system-level evaluation that sends the pest count after detection, acting as a live report within the final application scenario. 

The method by \textit{Suto}~\cite{sutHo2021embedded} reaches a detection accuracy of 82\%, similar to our solution, despite not being directly comparable because of the diverse dataset (not publicly available).
Their MCU, the STM32H7, consumes 30 J of energy for a MobileNetV2 inference, 6300× higher than our best solution executing a similar workload, 
{ Their MCU, the STM32H7, consumes \SI{30}{\joule} of energy for a MobileNetV2 inference, 6300$\times$ higher than our best solution executing a similar workload, leading to a daily consumption 6.9$\times$ higher than our CNN and a battery lifetime of only 4.6 days in the low-energy scenario. Furthermore, the platform requires 1 minute to perform the inference, which does not match the requirements of the high-frequency scenario. }

In contrast, the accuracy score (93\%) reported by~\cite{Brunelli_ours} only refers to the classification of a set of patches extracted from the frames of our same dataset. 
However, the background subtraction algorithm, which serves as a patch extractor, can lead to the misidentification of significant patches, thereby compromising the overall accuracy of the pest count. Variations in luminosity can result in the misclassification of these patches, a problem to which CNNs have demonstrated greater robustness~\cite{CNN_Luce}. Unlike background subtraction, which relies solely on detecting differences between frames, CNNs do not suffer from the inability to recover from false negatives. If a pest is not detected in the initial frame when it enters the trap, the background subtraction algorithm may subsequently classify it as part of the background, preventing its detection in future frames. In contrast, CNNs, which analyze each frame independently, are not subject to this limitation.
For these reasons, the detection accuracy of our method is not directly comparable with the score reported in~\cite{Brunelli_ours}. A fairer comparison between our methods is to apply their CNN, patch by patch, to the whole image, giving us an energy consumption of \SI{31.5}{\milli\joule} (6.6$\times$ more than ours).
Lastly, our object detection algorithm deployed on the GAP9 platform consumes \SI{4.85}{\milli\joule} of energy. 
This is only \SI{1.35}{\milli\joule} more than the energy consumption reported in~\cite{Brunelli_ours}, which used a low-frequency operating point (\SI{50}{\mega\hertz}) on the GAP8 platform. 
However, it is important to note that~\cite{Brunelli_ours} did not account for the energy cost of the external memories.

Since the system described in~\cite{Brunelli_ours} transmits image patches upon detection, its daily energy consumption is 1728.7 and \SI{1719.0}{\joule} for the high-frequency and low-energy scenarios, respectively. In this estimate, we considered the transmission cost from~\cite{Brunelli_ours} and an average of approximately 33 codling moths detected per trap per day, as reported in~\cite{croazia_cydia}. To fairly compare this system with our work, we calculated the energy consumption in the case our node sends an image after every detection (a total payload of \SI{12.7}{\kilo\byte} that includes a JPEG-compressed image and the overhead of the LoRaWAN protocol). Given that our LoRa module consumes \SI{1}{\milli\joule} per byte sent, the total energy consumption is \SI{12.7}{\joule} per image. For 33 detections per day, as reported in~\cite{croazia_cydia}, our system’s daily energy consumption is \SI{445.0}{\joule}  in the high-frequency scenario and \SI{423.4}{\joule} in the low-energy scenario. This daily energy consumption is approximately 3.9$\times$ lower than the system described in~\cite{Brunelli_ours} under the same conditions. Considering that we send only pest counters with the same frequency of detections, the energy consumption of our system is reduced by 6.5 times compared to sending the frame if a detection is made, achieving a lifetime of 199 days.

GAP9 achieved a 7$\times$ reduction in processing energy compared to our previous work~\cite{Ours_old}, which used GAP8 as the main processing unit. This significant reduction in processing cost decreases the daily energy consumption by \SI{76.2}{\joule}  for high-frequency and \SI{1.9}{\joule} for low-energy scenarios.
The proposed system can achieve a lifetime of 170 days for MBNV3-SSD and 199 days for Viola-Jones in the high-frequency scenario, and 2257 days for MBNV3-SSD and 2296 days for Viola-Jones in the low-energy scenario when sending pest counts at the same frequency as detections. In both scenarios, this exceeds the expected activity period of Cydia pomonella of approximately 120 days~\cite{Cidia_lifetime}.

\begin{table*}[t]
\caption{Comparison of Pest Detection Camera-based systems with ultra-low power MCUs for onboard processing.}
\label{tab:comparison}
\resizebox{1.0\textwidth}{!}{
\begin{threeparttable}
\begin{tabular}{|c|c|cccc|cccc|}
\hline
\multirow{3}{*}{}                                                     & \multirow{3}{*}{Method}                                                                                   & \multirow{3}{*}{Processor}                                                  & \multirow{3}{*}{\begin{tabular}[c]{@{}c@{}}Pest Detections\\ Latency\end{tabular}}                    & \multirow{3}{*}{\begin{tabular}[c]{@{}c@{}}Processing \\ Cost {[}\SI{}{\milli\joule}{]}\end{tabular}} & \multirow{3}{*}{\begin{tabular}[c]{@{}c@{}}Comm.\\ Module\end{tabular}}                              & \multicolumn{4}{c|}{Daily energy consumtpion {[}\SI{}{\joule}{]}}                                                                                                                                                                                                                                \\ \cline{7-10} 
                                                                      &                                                                                                           &                                                                             &                                                                                                       &                                                                                                       &                                                                                                      & \multicolumn{2}{c|}{High-frequency}                                                                                                                       & \multicolumn{2}{c|}{Low-energy}                                                                                                      \\ \cline{7-10} 
                                                                      &                                                                                                           &                                                                             &                                                                                                       &                                                                                                       &                                                                                                      & \begin{tabular}[c]{@{}c@{}}Transmitting\\ images\end{tabular} & \multicolumn{1}{c|}{\begin{tabular}[c]{@{}c@{}}Transmitting\\ pest counters\end{tabular}} & \begin{tabular}[c]{@{}c@{}}Transmitting\\ images\end{tabular} & \begin{tabular}[c]{@{}c@{}}Transmitting\\ pest counters\end{tabular} \\ \hline
J.Suto~\cite{sutHo2021embedded}                                                                & \begin{tabular}[c]{@{}c@{}}DNN\\  (Parameters: \SI{3.2}{\mega\nothing})\\ Acc:82\%\end{tabular}           & \begin{tabular}[c]{@{}c@{}}STM32H7 \\ \SI{480}{\mega\hertz}\end{tabular}    & \SI{1}{\minute}                                                                                       & 30600                                                                                                 & \begin{tabular}[c]{@{}c@{}}GSM\\ images\end{tabular}                                                 & ---                                                           & \multicolumn{1}{c|}{---}                                                                  & {2937.6}                                          & ---                                                                  \\ \hline
Brunelli et al.~\cite{Brunelli_ours}                                                       & \begin{tabular}[c]{@{}c@{}}GMM+DNN \\ (Parameters:ND)\\ Acc:93\%$^a$\end{tabular}                         & \begin{tabular}[c]{@{}c@{}}GAP8\\ \SI{50}{\mega\hertz}\end{tabular}         & \begin{tabular}[c]{@{}c@{}}\SI{810}{\milli\second} +\\  \SI{51}{\milli\second}/inference\end{tabular} & \textbf{3.5 $^b$}                                                                                     & \begin{tabular}[c]{@{}c@{}}LoRaWAN \\ images\end{tabular}                                            & {1728.7}                                          & \multicolumn{1}{c|}{---}                                                                  & {1719.0}                                          & ---                                                                  \\ \hline
Rusci et al.~\cite{Ours_old}                                                          & \begin{tabular}[c]{@{}c@{}}Viola-Jones \\ (Parameters: \SI{3.6}{\kilo\nothing})\\ Acc:81.7\%\end{tabular} & \begin{tabular}[c]{@{}c@{}}GAP8\\ \SI{175}{\mega\hertz}\end{tabular}        & \SI{400}{\milli\second}                                                                               & 31.8                                                                                                  & \multirow{3}{*}{\begin{tabular}[c]{@{}c@{}}\\LoRaWAN\\ pest \\ counter {\Big /}\\ images\end{tabular}} & {513.2}                                           & \multicolumn{1}{c|}{143.1}                                                                & {424.8}                                           & {7.7}                                                   \\ \cline{1-5} \cline{7-10} 
\multirow{2}{*}{\begin{tabular}[c]{@{}c@{}}This \\ Work\end{tabular}} & \begin{tabular}[c]{@{}c@{}}Viola-Jones\\ (Parameters: \SI{3.6}{\kilo\nothing})\\ Acc:81.7\%\end{tabular}  & \begin{tabular}[c]{@{}c@{}}GAP9 \\ \SI{240}{\mega\hertz}\end{tabular}       & \SI{221}{\milli\second}                                                                               & 4.61                                                                                                  &                                                                                                      & \textbf{{437.5}}                                  & \multicolumn{1}{c|}{\textbf{{66.9}}}                                          & \textbf{{423.3}}                                  & \textbf{{5.8}}                                           \\ \cline{2-5} \cline{7-10} 
                                                                      & \begin{tabular}[c]{@{}c@{}}MBNV3-SSD \\ (Parameters:\SI{3.44}{\mega\nothing})\\ Acc:83\%\end{tabular}     & \begin{tabular}[c]{@{}c@{}}GAP9 (NE16)\\ \SI{240}{\mega\hertz}\end{tabular} & \textbf{\SI{147}{\milli\second}}                                                                      & 4.85                                                                                                  &                                                                                                      & {445.0}                                           & \multicolumn{1}{c|}{{74.4}}                                                   & {423.4}                                           & {5.9}                                                    \\ \hline
\end{tabular}

\begin{tablenotes}
{
	 
	\item[{a}] only tested on crops,
	\item[{b}] not accounting for external memories,
    \
	}
\end{tablenotes}
\end{threeparttable}
}
\end{table*}
\section{Conclusion} \label{sec:conclusion}

In this work, we deployed a lightweight Viola-Jones and a CNN-based MobileNetV3-SSDLite algorithm for pest detection on a SoA multi-core MCU, the Greenwaves Technologies' GAP9, featuring a convolution hardware accelerator. We compare the performances of both methods in terms of efficiency and accuracy.
On two datasets, namely \textit{near-ds} and \textit{far-ds}, Viola-Jones achieved a detection accuracy of, respectively, 81.7\% and 45\%. 
The accuracy of this lightweight detector is outperformed by MBNV3-SSD in both datasets, scoring 83\% and 72\%, respectively.

\section*{Acknowledgments}
We thank Marco Fariselli, Tommaso Polonelli, and Carlo Montanari for their support. 
This work has been partially supported by the GEMINI ``Green Machine Learning for the IoT'' national research project, funded by the MUR under the PRIN 2022 program (Contract 20223M4HZ4).%
{
\section*{Conflict of Interest}
The authors of the manuscript declare no conflict of interest. Brand names and/or any mention or listing of specific commercial products or services herein are used solely for reference or comparison purposes. Such use does not imply endorsement by any of the manuscript's authors or affiliations.} 

\bstctlcite{IEEEexample:BSTcontrol}
\bibliographystyle{IEEEtran}
\bibliography{IEEEabrv,bibliography}

\newpage

\begin{IEEEbiography}[{\includegraphics[width=1in,height=1.25in,clip,keepaspectratio]{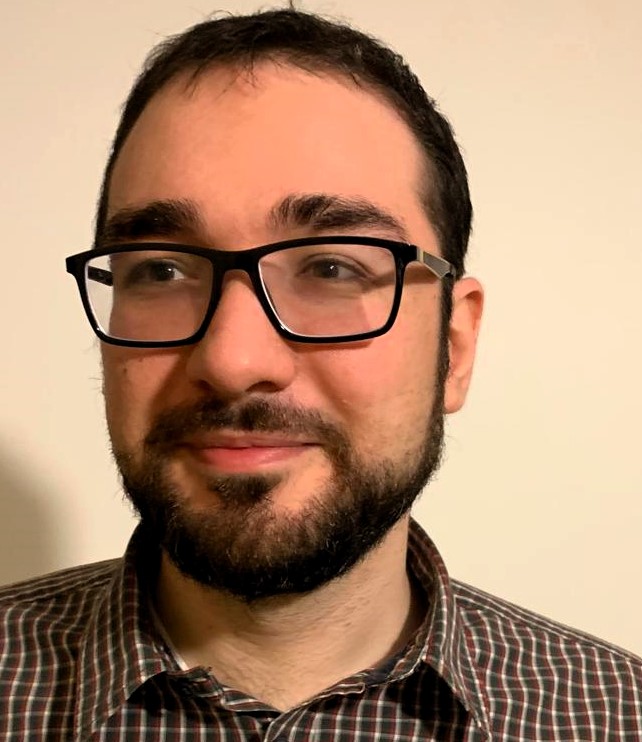}}]{Luca Bompani}(Student member, IEEE) 
currently a second-year Ph.D. student in Electronic Engineering at the University of Bologna. He received his MSc in theoretical physics and artificial intelligence at the University of Bologna. His research focuses on deploying and optimizing inference of Deep Neural Networks for energy-constrained ultra-low power embedded systems. He received the Best Paper Award at the IEEE/CVF CVPR'24 Embedded Vision workshop.
\end{IEEEbiography}

\begin{IEEEbiography}[{\includegraphics[width=1in,height=1.25in,clip,keepaspectratio]{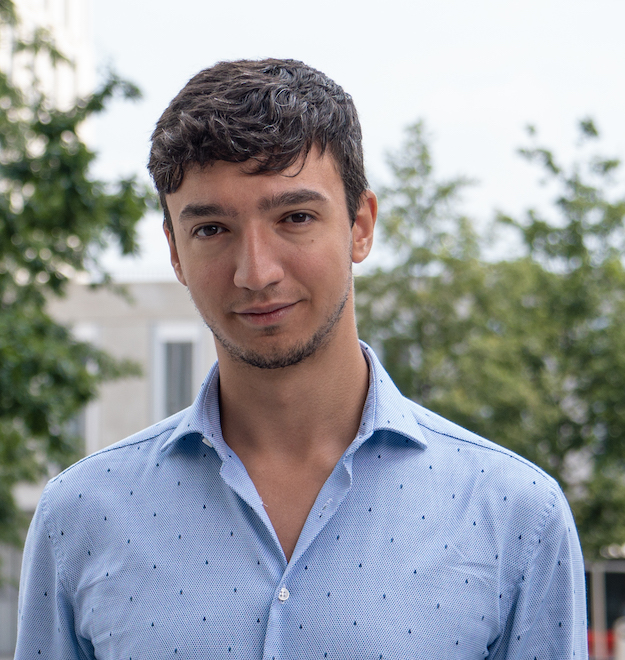}}]{Luca Crupi}(he/his) is a second-year Ph.D. student at the Dalle Molle Institute for Artificial Intelligence (IDSIA, USI-SUPSI) in Lugano, Switzerland. He was part of the XVII of Alta Scuola Politecnica and received an MSc degree in Computer Engineering from Politecnico di Torino in 2022. He was IT lead at PolitOcean, a student team focusing on the development of a fully autonomous underwater vehicle. His research focuses on developing AI-based autonomous algorithms for pocket-sized robotic platforms integrating multiple sensor streams to perform sensing with neural network-based techniques. His work has resulted in 8 peer-reviewed publications in international conferences and journals.
\end{IEEEbiography}

\begin{IEEEbiography}[{\includegraphics[width=1in,height=1.25in,clip,keepaspectratio]{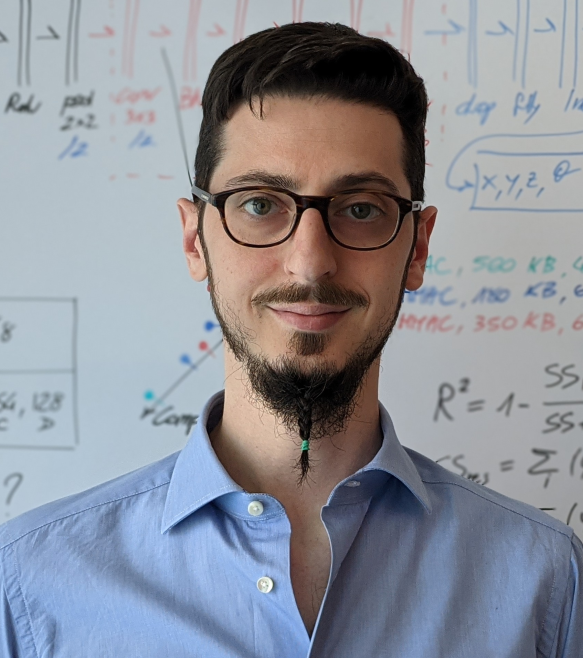}}]{Daniele Palossi} (he/his) received his Ph.D. in Information Technology and Electrical Engineering from ETH Z\"urich. He is currently a Senior Researcher at the Dalle Molle Institute for Artificial Intelligence (IDSIA), USI-SUPSI, Lugano, Switzerland, where he leads the nano-robotics research group, and at the Integrated Systems Laboratory (IIS), ETH Z\"urich, Z\"urich, Switzerland. His research stands at the intersection of artificial intelligence, ultra-low-power embedded systems, and miniaturized robotics. His work has resulted in 45+ peer-reviewed publications in international conferences and journals. Dr. Palossi was a recipient of the Swiss National Science Foundation (SNSF) Spark Grant, the 2nd prize at the Design Contest held at the ACM/IEEE ISLPED'19, several Best Paper Awards, and team leader of the winning team of the first ``Nanocopter AI Challenge'' hosted at the IMAV'22 International Conference.
\end{IEEEbiography}

\begin{IEEEbiography}[{\includegraphics[width=1in,height=1.25in,clip,keepaspectratio]{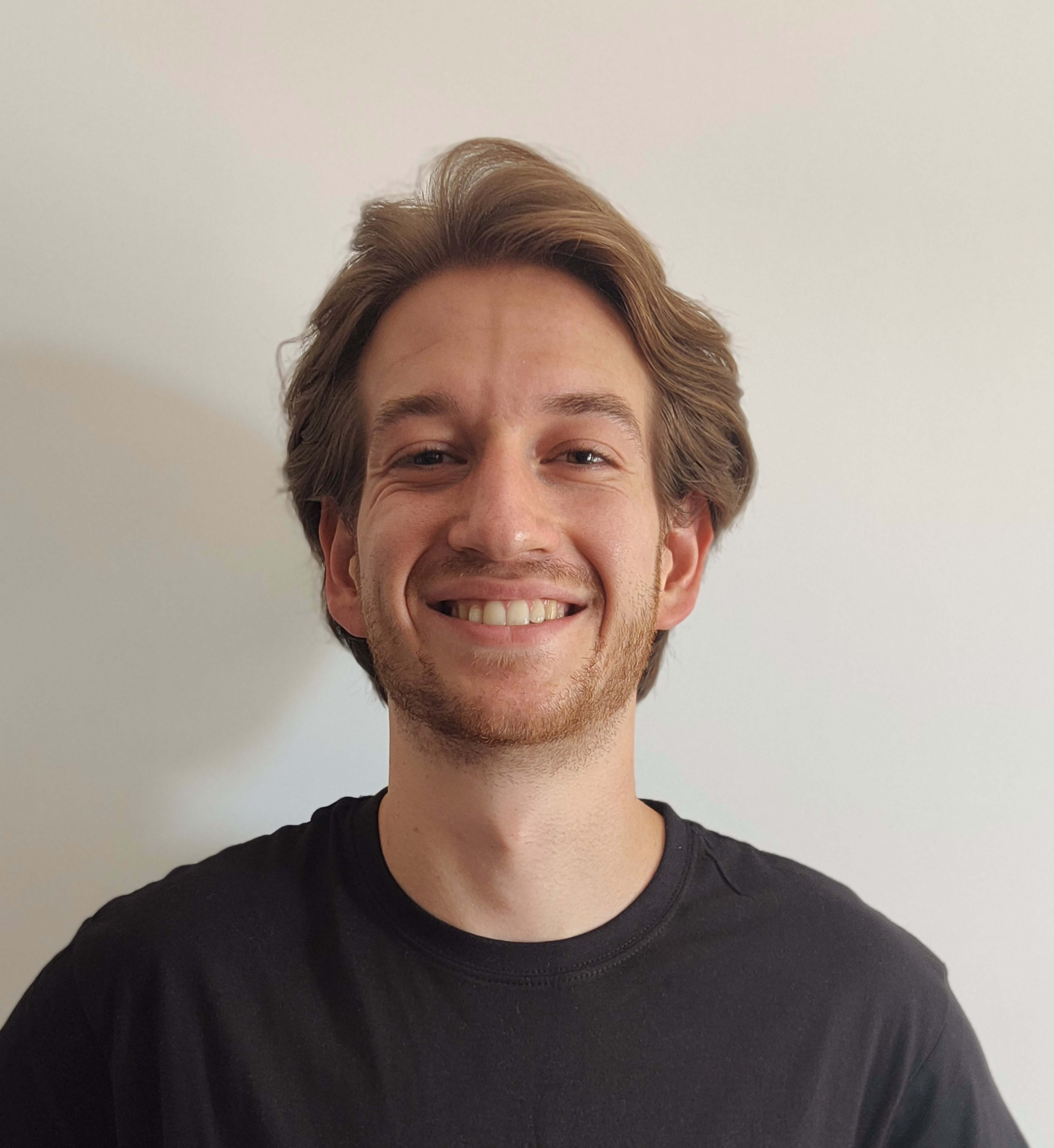}}]{Olmo Baldoni}is a computer engineering student with a deep passion for machine learning and deep learning. After graduating in Automation Engineering at the University of Bologna, he continued his studies with a Master's degree in Computer Engineering at the University of Modena and Reggio Emilia. \end{IEEEbiography}

\begin{IEEEbiography}[{\includegraphics[width=1in,height=1.25in,clip,keepaspectratio]{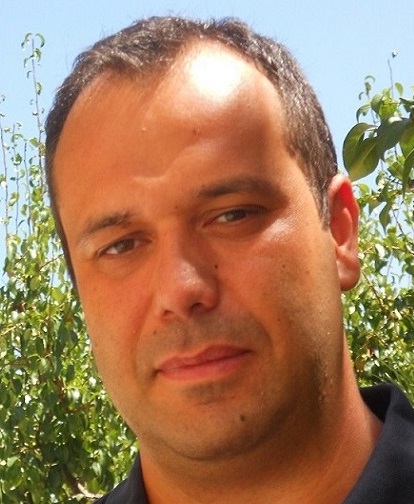}}]{Davide Brunelli} (Senior Member, IEEE) received the M.S. (cum laude) and Ph.D. degrees in electrical engineering from the University of Bologna, Italy, in 2002 and 2007, respectively. He is currently an Associate Professor of electronics with the Department of Industrial Engineering, University of Trento, Italy. He has authored or coauthored more than 280 research papers in international conferences and journals on ultralow-power embedded systems, energy harvesting, and power management of VLSI circuits. He holds several patents and is annually ranked among the top 2\% of scientists according to the “Stanford World Ranking of Scientists” from 2020. His research interests include new techniques of energy scavenging for IoT and embedded systems, the optimization of low-power and low-cost consumer electronics, and the interaction and design issues in embedded personal and wearable devices. Prof. Brunelli is a Member of several TPC conferences on the Internet of Things (IoT) and is an Associate Editor for the IEEE TRANSACTIONS ON AGRIFOOD ELECTRONICS.\end{IEEEbiography}

\begin{IEEEbiography}[{\includegraphics[width=1in,height=1.25in,clip,keepaspectratio]{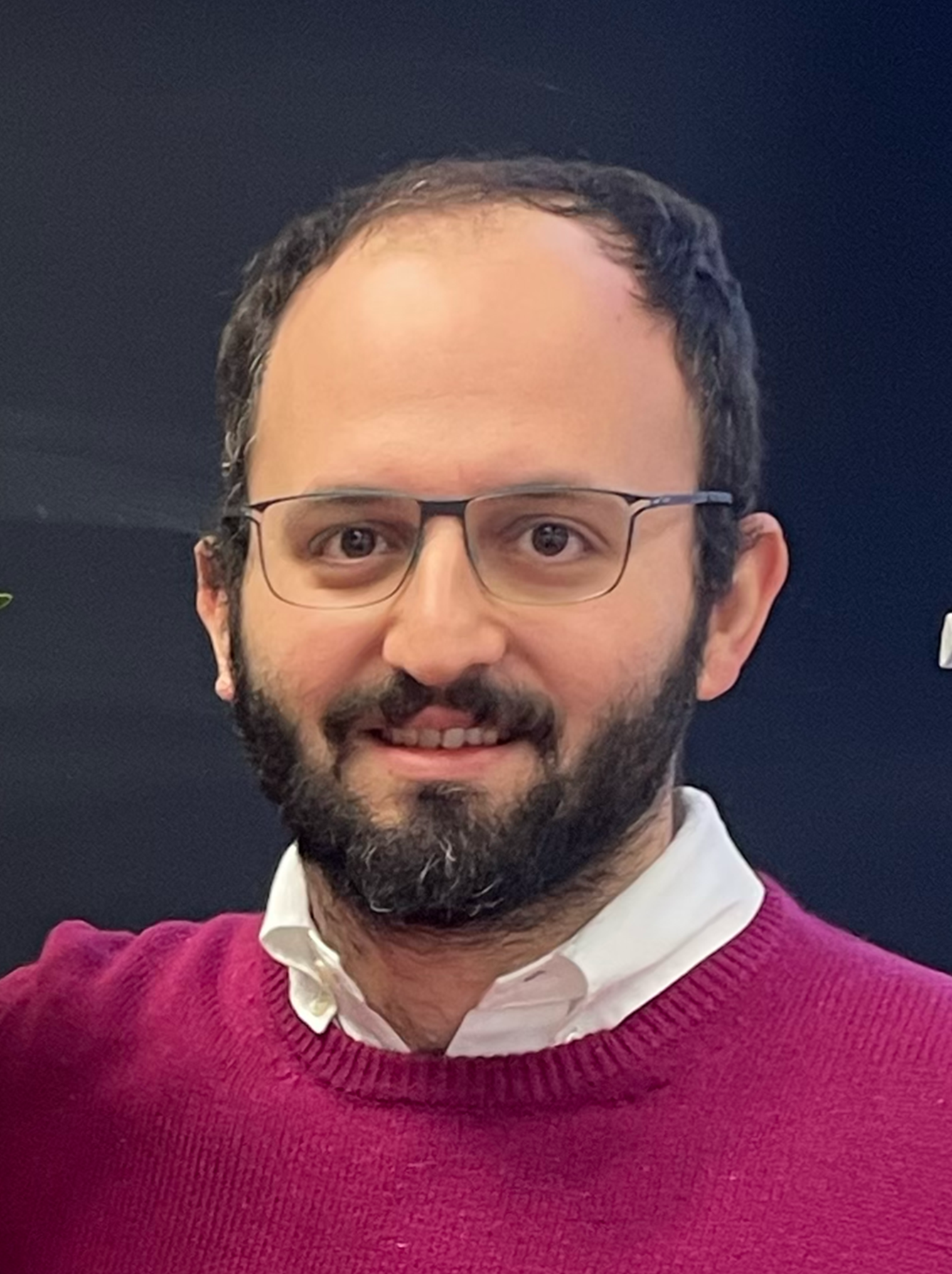}}]{Francesco Conti} (Member, IEEE) received his Ph.D. degree in electronic engineering from the University of Bologna, Italy, in 2016. He is currently a Tenure-Track Assistant Professor with the DEI Department at the University of Bologna. From 2016 to 2020, he held a research grant with the University of Bologna and a position as a Post-Doctoral Researcher with ETH Z\"urich. His research is centered on hardware acceleration in ultra-low power and highly energy-efficient platforms, with a particular focus on System-on-Chips for Artificial Intelligence applications. His research work has resulted in more than 100 publications in international conferences and journals and was awarded several times, including the 2020 IEEE \textsc{Transactions on Circuits and Systems I: Regular Papers} Darlington Best Paper Award.
\end{IEEEbiography}

\begin{IEEEbiography}[{\includegraphics[width=1in,height=1.25in,clip,keepaspectratio]{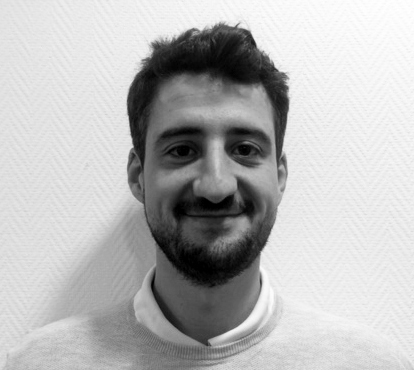}}]{Manuele Rusci}
received the Ph.D. degree in electronic engineering from the University of
Bologna in 2018. He is currently holding a MSCA Post-Doctoral Fellowship at the Katholieke Universiteit Leuven, after being Post-Doc at the University of Bologna. His main research interests include low-power AI-powered smart sensors and on-device continual learning. 
\end{IEEEbiography}

\begin{IEEEbiography}[{\includegraphics[width=1in,height=1.25in,clip,keepaspectratio]{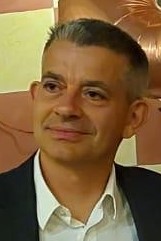}}]{Luca Benini}holds the chair of digital Circuits and systems at ETHZ and is Full Professor at the Università di Bologna. He received a PhD from Stanford University.  His research interests are in energy-efficient parallel computing systems, smart sensing micro-systems and machine learning hardware. He is a Fellow of the ACM, a member of the Academia Europaea and of the Italian Academy of Engineering and Technology.
\end{IEEEbiography}

\end{document}